# Investigation of channel model for weakly coupled multicore fiber

LIN GAN,1 LI SHEN,1 MING TANG1,* CHEN XING,1 YANPENG LI,1 CHANGJIAN KE,1 WEIJUN TONG,2 BORUI LI,1 SONGNIAN FU,1 AND DEMING LIU,1

1 Wuhan National Lab for Optoelectronics (WNLO) & National Engineering Laboratory for Next Generation Internet Access System, School of Optics and Electronic Information, Huazhong University of Science and Technology
2 State Key Laboratory of Optical Fiber and Cable Manufacture Technology, Yangtze Optical Fiber and Cable Joint Stock Limited Company (YOFC). R&D Center
*tangming@mail.hust.edu.cn

Abstract:
   We investigate the evolution of decorrelation bandwidth of inter core crosstalk (IC-XT) in homogeneous weakly coupled multicore fibers (WC-MCFs). The modified mode-coupled equations (MCEs) are numerically solved by combining the fourth order Runge-Kutta method and compound Simpson integral method. It can be theoretically and numerically observed that the decorrelation bandwidth of IC-XT decreases with transmission distance by fractional linear function. The evolution rule of IC-XT's decorrelation bandwidth is further confirmed by experiments, which can be used as an evaluation criterion for channel model. Finally, we propose a new channel model with the coupling matrix of IC-XT generated automatically by phase transfer function (PTF), which is in good agreement with the above evaluation criterion. We believe the proposed channel model can provide a good simulation platform for homogeneous WC-MCF based communication systems.

1. Introduction

   Space division multiplexing (SDM) technologies are introduced to improve the capacity about ten to hundred times compared with the communication systems based on single mode fibers (SMFs) [1-3]. Several kinds of SDM fibers have been proposed, such as weakly coupled multicore fiber (WC-MCF) [4, 5], strongly coupled multicore fiber (SC-MCF)[6-8], few mode fiber (FMF) [9-11], and few mode multicore fiber (FM-MCF) [12, 13]. All the cores and modes can be called spatial channels. And the random power coupling between spatial channels can be called crosstalk (XT). On one hand, there are no need for multiple-input multiple-output (MIMO) algorithms to compensate XT in WC-MCF due to ultra-low XT [14, 15]. However, the MIMO algorithms are necessary for SC-MCF [7, 8, 16], FMF(not specially designed, such as elliptical core) [9-11] and FM-MCF [13] due to unneglectable XT, which will increase system complexity. On the other hand, taking into account the spatial efficiency (defined as the spatial multiplicity divided by the fiber cross-sectional area), WC-MCF has lower spatial efficiency than SC-MCF, FMF and FM-MCF [17], which will decrease the system capacity. To figure out which type is the best SDM fiber in different transmission scenarios, one way is to compare their transmission performance when using different types of SDM fibers. Combining with other factors, such as system complexity, energy consumption, and hardware cost, we can choose the best type of SDM fibers and design the SDM fiber structure more suitable for different scenarios .

   To evaluate the performance of SMFs based communication system, SMF's channel model has been proposed. Propagation effects of optical signals can be modeled effectively and precisely based on the nonlinear Schrodinger equation (NLSE) [18]. For SDM fibers, propagation effects in each spatial channel can also be described by the NLSE when there is ignorable XT between spatial channels. However, in general, the XT cannot be ignored. Therefore, to accurately evaluate the performance of SDM fiber based systems, we need to investigate the XT's effects on signal quality. Firstly, we need XT model, which is mainly focused on analyzing average XT power evolution, time-dependent characteristics, frequency-dependent characteristics and other characteristics of XT caused by different fiber structures and external conditions, such as input optical power, bending, twist and other fluctuations. Further, the channel model of SDM fibers can be constructed based on XT model, which is focused on analyzing the transmission effects (especially the XT) on optical signals.

For FMF, the XT can be defined as the random power coupling between different fiber modes. The FMF's XT model shows that the coupling strength between any two modes strongly depends on the difference of propagation constant between the two modes [19]. Therefore, the modes that have similar propagation constants couple strongly. And it is always assumed that the mode coupling of different frequencies are the same [20]. The FMF's channel model has been proposed by using the coupled NLSEs based on related conclusions of FMF's XT model [21-23]. A differential operator can describe mode dependent loss (MDL), differential group delay (DGD) and group-velocity dispersion (GVD). And the effects of random mode coupling within a group are modeled by the generalized Manakov equations with polarization mode dispersion (PMD) taking into consideration [21, 24]. In addition, the intergroup coupling is modeled using the exponent of the random Hermitian matrix [25, 26]. The coupled NLSEs are always numerically solved by split-step Fourier method [18]. The existing FMF's channel model can well describe the propagation effects of optical signals in FMFs, which has been demonstrated in VPItransmissionMaker Optical Systems 9.8.

For WC-MCF's XT model, the XT can be defined as the random power coupling between different cores. The inter core crosstalk (IC-XT) can be modeled by the modified mode coupled equations (MCEs) [4, 27]. Further, for homogeneous WC-MCF, when the fiber operates in the phase-matching region (when the bending radius is under the threshold bending radius), the average IC-XT power mainly changes at each phase matching points (PMPs) [4, 28]. In addition, the PMPs are sensitive to the propagation constant mismatch, bending radius, twist speed and other fluctuations [28]. The average IC-XT power has been proved changing linearly with transmission distance, and bending radius with the assumption of ultra-low IC-XT [4, 28]. In addition, the linear and nonlinear IC-XT power has been investigated by using the nonlinear MCEs [29]. For heterogeneous WC-MCF, the average IC-XT power evolution has been analyzed using the modified MCEs [27]. Recently, for frequency-dependent characteristics of IC-XT [30] in homogeneous WC-MCF, with the assumption of ultra-low IC-XT, the effects of group velocity's difference and GVD's difference on crosstalk transfer function (XTTF) have been discussed based on a novel analytical IC-XT model [31, 32]. The model has been further improved into an analytical discrete changes model [33, 34], which is suitable for analyzing the frequency-dependent IC-XT in real homogeneous WC-MCF when the fiber operates in phase-matching region. In addition, the time-dependent characteristics of frequency-dependent IC-XT have been investigated with different signal bandwidth and formats [35, 36].

For WC-MCF's channel model, similar to the FMF's channel model, the linear and nonlinear propagation effects can also be modeled using coupled NLSEs with IC-XT described discretely using the coupling matrix [21, 37]. However, there are two strategies to construct the IC-XT's coupling matrix for different signal bandwidths in homogeneous WC-MCF. As described in [38-40], for adequately narrow band signals, the IC-XT can be described as multiple delayed copies of the signals with uncorrelated phase. And for the adequately broad band signals, the IC-XT can be described as a virtual additive white Gaussian noise (AWGN) on I-Q planes due to the IC-XT changing rapidly with optical frequency [30, 39]. To unify the two strategies, in our previous work [41, 42], we modeled the IC-XT in frequency domain. The spectrum of optical signals is manually divided into multiple frequency segments with different IC-XT's coupling matrix. For optical signals within one frequency segment, the IC-XT is treated as signal copies with random phase according to the first strategy. Due to the IC-XT changing rapidly with optical frequency, we set the random phase of different frequency segments generated independently according to the second strategy. In [39], the IC-XT spectrum of a 17 km homogeneous WC-MCF has been measured. The decorrelation bandwidth (half width at 1/e) calculated from the autocorrelation function of the spectrum is about 10-20 pm (~1-2GHz). Therefore, we manually divide the frequency segments with the width of each frequency segment equivalent to the decorrelation bandwidth. However, there are two important issues which need be further investigated. The first issue is whether the decorrelation bandwidth (the width of each frequency segment) should be always about 1~2 GHz with the transmission distance. The second is how to construct the IC-XT's coupling matrix automatically instead of manually.

  On the other hand, for SC-MCF, defined as MCF with a Gaussian-like impulse response, the core pitch of SC-MCF is always around 20-25 um [16]. The IC-XT of SC-MCF are also can be modeled using modified MCEs [16, 29], which is the same as WC-MCF. When the core pitch is less than 20 um, the MCF need to be described using super-modes rather than the modified MCEs [43]. Therefore, we do not discuss the super-mode MCF in this paper. For SC-MCF's channel model, the propagation effects can also be modeled using coupled NLSEs with IC-XT described discretely using the coupling matrix due to the similarity between WC-MCF and SC-MCF. As proposed in [16], the IC-XT of SC-MCF's channel model is treated as signal copies without frequency-dependent characteristics.

  Therefore, for FMF, the channel model can well model the propagation effects. But for WC-MCF and SC-MCF, the channel model need to be further discussed. For frequency-dependent characteristics of IC-XT, the channel model remains some issues to be solved as mentioned above. Although both the IC-XT in WC-MCF and SC-MCF can be modeled by the modified MCEs, the existing XT model can only suitable for analyzing the IC-XT's frequency-dependent characteristics in homogeneous WC-MCF when the fiber operates in phase-matching region. The reason is the basic expression of frequency-dependent IC-XT needs the assumption of ultra-low XT and PMPs.

  In this paper, to avoid the assumptions of ultra-low XT and PMPs, we choose to combine the fourth order Runge-Kutta method and compound Simpson integral method to directly solve the modified MCEs. The proposed numerical simulation method can suitable for analyzing the modified MCEs in heterogeneous or homogeneous WC- or SC- MCF. Limited by the experimental conditions, we only verified the results of homogeneous WC-MCF in phase-matching region. To figure out the decorrelation bandwidth's evolution with transmission distance, we numerically solve the modified MCEs with different optical frequencies under the same condition. It can be observed theoretically and numerically that the decorrelation bandwidth (half width at 1 dB) of XT decreases with transmission distance by fractional linear function, which is further confirmed by experimental results. Thus, we can use the evolution rule of XT model as an evaluation criterion for homogeneous WC-MCF's channel model. To construct the coupling matrix automatically, we propose a new channel model based on phase transfer function (PTF), which is suitable for homogeneous WC-MCF in phase-matching region. The XT's frequency-dependent characteristics of the new MCF's channel model are confirmed by the above evaluation criterion. We believe the proposed channel model can provide a good simulation platform for homogeneous WC-MCF based communication systems.

2. Numerical simulation of IC-XT

2.1. Review of the existing XT model for homogeneous WC-MCF

Here we briefly review the details of the existing XT model for homogeneous WC-MCF based on the modified MCEs [4]. For homogeneous WC-MCF, the major external factors affecting IC-XT are fiber bending and twist, and the major internal factors are core pitch and the refractive index (RI) difference [28, 44]. When assume the fiber is lossless, the modified MCEs are shown as Eq. (1), where the symbol A and B represent the electric fields in Core A and Core B, respectively. $z$ is transmission distance. $\lambda$ is optical wavelength. $\kappa(\lambda)$ represents the coupling coefficient between Core A and Core B.

$$\begin{aligned}\frac{dA}{dz} &= -j\kappa(\lambda)B\exp(-j\Delta\phi(z,\lambda)) \\ \frac{dB}{dz} &= -j\kappa(\lambda)A\exp(+j\Delta\phi(z,\lambda))\end{aligned} \quad (1)$$

In Eq. (1), $\Delta\phi(z,\lambda)$ is the relative phase mismatch (RPM) between Core A and Core B, which can be represented as Eq. (2).

$$\Delta\phi(z,\lambda) = \int_0^z [\beta_{eq,B}(z',\lambda) - \beta_{eq,A}(z',\lambda)]dz' \quad (2)$$

where $\beta_{eq,A}$ and $\beta_{eq,B}$ are the equivalent propagation constants of Core A and Core B. The equivalent propagation constant $\beta_{eq,n}$ of Core n is represented as Eq. (3) based on the equivalent refractive index (ERI) model [44].

$$\beta_{eq,n}(z,\lambda) = \beta_{c,n}(\lambda)\cdot\left(1+r_n\frac{\cos\theta_n(z)}{R_b(z)}\right) \quad (3)$$

where $\beta_{c,n}(\lambda)$ is the unperturbed propagation constant $\beta_{c,n}(\lambda) = k(\lambda)\cdot n_{eff,n}(\lambda)$, $k(\lambda) = 2\pi/\lambda$ is the wave number in a vacuum, $n_{eff}(\lambda)$ is the effective refractive index of fundamental mode. As shown in Fig. 1, $R_b$ represents the bending radius of MCF, and $(r,\theta)$ represents the angle of the local polar coordinate in the cross-section of the MCF with $\theta=0$ in the radial direction of the bend. We assume Core A is the center core, and Core B is the side core.

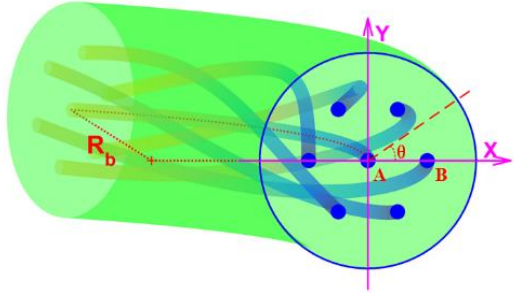

Fig. 1. Schematic diagram of seven core homogeneous WC-MCF.

When the fiber is bending and without twist, the IC-XT is ultra-low because of the mismatch of propagation constants between Core A and Core B, which is mainly caused by the inherent RI mismatch due to the actual production of fiber preform and the ERI mismatch due to the fiber bending [28]. When the fiber twists, the mismatch of the propagation constants would be zero at some points, which are called PMP. Since $\beta_{eq,A}$ and $\beta_{eq,B}$ are equal at each PMP, it will introduce a strong power coupling between Core A and Core B. For WC-MCF, we assume the electric field amplitudes of Core A and Core B are 1.0 and 0.0 at start point $z=0$. Thus, the Core A is interfering core and Core B is interfered core. With the assumption of ultra-low XT and the unchanged amplitude of Core A, the IC-XT in Core B can be represented as [4]

$$B_N(z) = B_0 - j|K|\sum_{k=1}^{N}\exp[-j\phi_{rnd}(k)]A_{k-1}$$
$$= -j|K|\sum_{k=1}^{N}\exp[-j\phi_{rnd}(k)] \quad (4)$$

where $B_N$ is the amplitude of Core B at Nth PMP, $\phi_{rnd}$ represents the random phase mismatch which is always assumed uniformly distributed $[0, 2\pi]$. And

$$|K| \cong \sqrt{\frac{\kappa^2}{\beta_c}\frac{R_b}{D}\frac{2\pi}{\gamma}} \quad (5)$$

where $D$ is the distance between Core A and Core B. $\gamma$ is the twist speed. According to the central limit theorem, the real part and imaginary part of electric filed amplitude of Core B are Gaussian distributed whose variance $\sigma^2$ is $N|K|^2/2$ if $N$ is adequately large. Thus, the average IC-XT power $XT_{ave}$ can be obtained as Eq. (6)

$$XT_{ave} = 2\sigma_2^2 = 4\sigma_4^2 = 2\frac{\kappa^2}{\beta}\frac{R_b}{D}L = hL \quad (6)$$

where $\sigma_2^2$ is for single polarization modes and $\sigma_4^2$ is for two polarization modes. $L$ is transmission distance. As shown in Eq. (6), the average IC-XT power will change linearly with bending radius and transmission length in phase-matching region.

For frequency-dependent characteristics of IC-XT in homogeneous WC-MCF, it is always assumed that the XT power is very low and the amplitude of Core A is unchanged. Therefore, the expression of IC-XT (Eq. (4)) in Core B can be rewritten as Eq. (7) when the MCF operates in phase-matching region [31-33].

$$XT(\omega) = -j|K|\exp\left(-j\beta_{eq,B}(\omega)L\right)\sum_{k=1}^{N}\exp\left(-j\Delta\beta_{eq}(\omega)z_k - j\phi_{rnd}(k)\right) \quad (7)$$

The value $K$ has been further investigated in [33]. The difference of propagation constants $\Delta\beta_{eq}(\omega)$ can be written in Taylor series with optical frequency $\omega$ represented as Eq. (8) [18].

$$\Delta\beta_{eq}(\omega) = \beta_{eq,B}(\omega) - \beta_{eq,A}(\omega)$$
$$= (\beta_{0,B} - \beta_{0,A})\omega_0 + (\beta_{1,B} - \beta_{1,A})(\omega-\omega_0) + \frac{1}{2}(\beta_{2,B} - \beta_{2,A})(\omega-\omega_0)^2 + ... \quad (8)$$
$$= \Delta\beta_0\omega_0 + \Delta\beta_1(\omega-\omega_0) + \frac{1}{2}\Delta\beta_2(\omega-\omega_0)^2 + ...$$

where $\omega_0$ is the center optical frequency. $\beta_{0,n}$, $\beta_{1,n}$, and $\beta_{2,n}$ are the Taylor series of equivalent propagation constants $\beta_{eq,n}$ of Core n. Base on the Eq. (7) and Eq. (8), the effects of group velocity walk off and GVD on frequency-dependent IC-XT has been investigated [31, 32].

It should be noted that the existing XT model can only suitable for analyzing the IC-XT's frequency-dependent characteristics in homogeneous WC-MCF when the fiber operates in phase-matching region. The reason is the basic expression of frequency-dependent IC-XT needs the assumption of ultra-low XT and PMPs.

## 2.2. New numerical simulation method to solve the modified MCEs

To avoid the assumptions of ultra-low XT and PMPs [31-33], one way is to numerically solve the modified MCEs (Eq. (1)) directly. Without these assumptions, the XT in heterogeneous or homogeneous WC- or SC-MCF can be analyzed simultaneously. Since the Core A is the center core, and Core B is the side core, we rewrite the Eq. (3) with frequency-dependent characteristics taking into consideration, as shown in Eq. (9) and Eq. (10).

$$\beta_{eq,A}(z,\omega) = \beta_{c,A}(\omega) \cdot (1 + S_{\beta,A}(z)) \quad (9)$$

$$\beta_{eq,B}(z,\omega) = \beta_{c,B}(\omega) \cdot (1 + Bias_\beta + S_{\beta,B}(z)) \cdot \left(1 + D\frac{\cos\theta(z)}{R_b(z)}\right) \quad (10)$$

$S_{\beta,A}$ and $S_{\beta,B}$ are the longitudinal fluctuations of propagation constants of Core A and Core B, respectively, which are caused by inherent fluctuations, such as structure fluctuations, fabrication fluctuations. $S_{\beta,A}$ and $S_{\beta,B}$ are generated independently with each other and uniformly distributed with interval $[-S_A/2, S_A/2]$ and $[-S_B/2, S_B/2]$, respectively. $Bias_\beta$ is the inherent propagation constant mismatch caused by fabrication error which is related to the threshold bending radius $R_{pk}$ demonstrated in [28]. We assume $Bias_\beta$ is constant with distance. And we also assume that both the core pitch $D$ and the coupling coefficient $\kappa$ are constant with distance, because of the good geometric consistency of fabricated MCF. The unperturbed propagation constant $\beta_c$ is generated by numerically solving the dispersion equation (Eq. 3.40) in [45]. For homogeneous WC-MCF and SC-MCF, all the cores share the same designed structures, and the $\beta_{c,A}$ is equal to $\beta_{c,B}$ with weak propagation constants mismatch $Bias_\beta$. For heterogeneous WC-MCF and SC-MCF, the design parameters of different cores are different. Thus, the $\beta_{c,A}$ is not equal to $\beta_{c,B}$. The coupling coefficient $\kappa$ is the numerical integration (Eq. 4.13) in [45], which can be verified with finite element method (FEM) shown in [46].

The bending radius $R_b(z)$ and local angle $\theta(z)$ of Core B can be represented as Eq. (11) and Eq. (12), respectively. The variations of bending radius $S_R$ and twist speed $S_T$ are also assumed uniformly distributed with interval $[-S_{RB}/2, S_{RB}/2]$ and $[-S_{TB}/2, S_{TB}/2]$, respectively. The distribution type of the variations $S_{\beta,A}$, $S_{\beta,B}$, $S_R$, $S_T$ should have little influence on simulation results, because the mainly purpose is to induce the random phase mismatch $\phi_{rnd}$ in Eq. (4) and Eq. (7), which is always assumed uniformly distributed within $[0, 2\pi]$.

$$R_b(z) = R_0(1 + S_R(z)) \quad (11)$$

$$\theta(z) = \int_0^z \{\gamma[1 + S_T(z')]z' + \theta_0\}dz' \quad (12)$$

Based on the above assumptions, we can combine the fourth order Runge-Kutta method and compound Simpson integral method to solve the modified MCEs (Eq. (1)). In details, the differential equations of MCEs is solved by fourth order Runge-Kutta method [47], and the RPM (Eq. (2)) is solved by compound Simpson integral method [47]. It should be noted that the calculation step size of compound Simpson integral method is half of the calculation step size of the fourth order Runge-Kutta method in order to ensure the accuracy. However, the integration of Eq. (12) is accumulated with distance directly to decrease the computation complexity. The decorrelation lengths of $S_{\beta,A}$, $S_{\beta,B}$, $S_R$, $S_T$ are all 0.1 m in the followed numerical simulation. Compared to the calculation step size (1.0 um to 1.0 cm) and simulation total length (100.0-500.0 m), the correlation length from 1.0 cm to 1.0 m should be accepted to generate randomness. And we also set the initial angle $\theta_0$ to zero in the followed numerical simulation. The time consumption of each calculation step is around 3.5 microseconds based on CPU Intel i5-2450M and MATLAB 2017a.

Based on the above numerical calculation method, we can analyze the evolution of XT in homogeneous or heterogeneous WC-MCF and SC-MCF with arbitrary bending radius and optical frequency. Due to the experimental conditions, we only focus on the homogeneous WC-MCF in phase-matching region with step refractive index profile operating in single mode. The core diameter, the core pitch, the RI of each core and the RI of cladding are 8.7 um, 41.1 um, 1.4639 and 1.4591, respectively.

2.3. Verify the accuracy of numerical simulation

Firstly, we should verify the calculation accuracy compared with the analytical results of directional straight coupler described in Chapter 4 of [45]. Therefore, the bending radius and twist speed are infinite and zero, respectively. the $Bias_\beta$, $S_A$, $S_B$, $S_{RB}$, $S_{TB}$ are all zero. The optical wavelength, the calculation step size and transmission distance are 1550.4 nm, 1.0 um and 500.0 m, respectively. Fig. 1(a) shows the numerical simulation results. The analytical results can be represented as Eq. (13) where $A(0) = 1.0$ and $B(0) = 0.0$. $z$ is the transmission distance. $\kappa$ is the coupling coefficient calculated by the numerical integration (Eq. 4.13) in [45]. The maximum coupling power of numerical simulation and analytical results are all 1.0 because there is no propagation constant mismatch.

$$\begin{bmatrix} A(z) \\ B(z) \end{bmatrix} = \begin{bmatrix} \cos(\kappa z) & -j\sin(\kappa z) \\ -j\sin(\kappa z) & \cos(\kappa z) \end{bmatrix} \begin{bmatrix} A(0) \\ B(0) \end{bmatrix} \qquad (13)$$

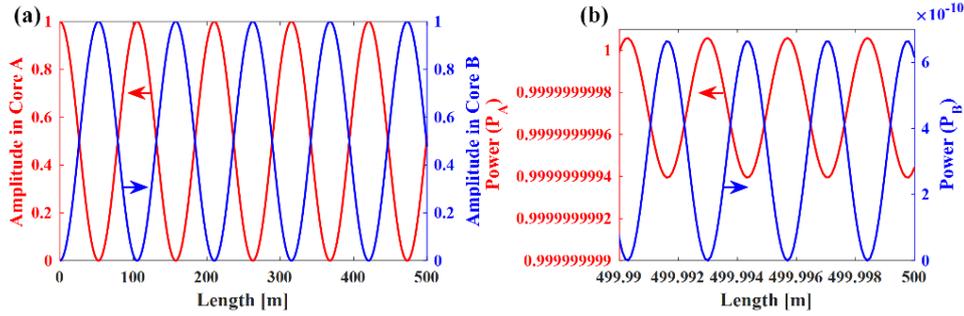

Fig. 2. (a)Numerical results without bending. (b)Numerical results with bending radius 105 mm.

Secondly, to verify the simulation with propagation constant mismatch, we set the bending radius 105 mm with other parameters unchanged. Fig. 2(b) shows the numerical simulation results. The maximum coupling power and coupling period are $6.64863940 \times 10^{-10}$ and 2.71 mm, respectively. The analytical results can be represented as Eq. (14) [45]. $\beta_{eq,A}$ and $\beta_{eq,B}$ are calculated by Eq. (9) and Eq. (10). The $\beta_c$ is generated by numerically solving the dispersion equation (Eq. 3.40) in [45]. The analytical maximum coupling power and coupling period are $6.64863931 \times 10^{-10}$ and 2.71002 mm, respectively. It can be concluded that the accuracy of numerical simulation is very high.

$$B(z) = -j\frac{\kappa}{q}\sin(qz)\exp(j\delta z); where\ q = \sqrt{\kappa^2 + \delta^2}, \delta = \frac{\beta_{eq,B} - \beta_{eq,A}}{2} \qquad (14)$$

In order to increase the calculation step size for saving time while guaranteeing the accuracy, we calculate the XT under the condition shown in Fig. 2(b) with calculation step size changing from $1.0 \times 10^{-6}$ m to $1.0 \times 10^{-2}$ m. The relative error at 500.0 m can be calculated by comparing numerical simulation results with the analytical results of Eq. (14). As shown in Fig. (3). the best calculation step size should be $1.0 \times 10^{-4}$ m.

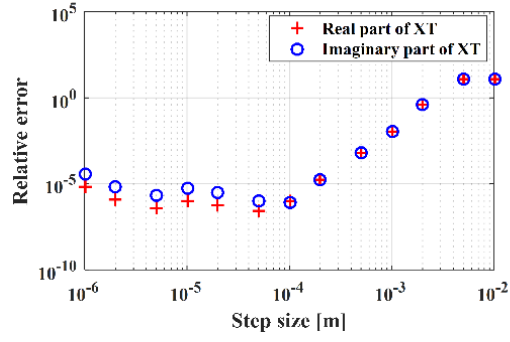

Fig. 3. Relative error of numerical simulation under different calculation step sizes.

2.4. Verify the numerical simulation suitable for existing XT model

To verify the numerical simulation method, we will discuss the XT under three different conditions. Under the first condition, the bending radius and twist speed are 105 mm and 4.0 rounds per meter, respectively. the $Bias_\beta$, $S_A$, $S_B$, $S_{RB}$, $S_{TB}$ are all zero. The optical wavelength, the calculation step size and transmission total length are 1550.4 nm, $1.0 \times 10^{-4}$ m and 500.0 m, respectively. Due to unavoidable fabrication defects, the refractive index of different cores in homogenous WC-MCF are slightly different, which will cause the propagation constant mismatch $Bias_\beta$ [28]. Therefore, under the second condition, the $Bias_\beta$ is $2.0 \times 10^{-5}$ with other parameters unchanged compared to the first condition. To simulate a real homogenous WC-MCF in real environment, under the third condition, we add reasonable fluctuations [44]. The $S_A$, $S_B$, $S_{RB}$, $S_{TB}$ are $5.0 \times 10^{-5}$, $5.0 \times 10^{-5}$, $1.0 \times 10^{-2}$, $1.0 \times 10^{-2}$, respectively, with other parameters unchanged compared to the second condition. The $S_A$, $S_B$, $S_{RB}$, $S_{TB}$ are mainly used to generate random phase of Eq. (2). The XT's characteristics are insensitive to the specific value of those parameters as long as the RPM is sufficiently random at each PMP.

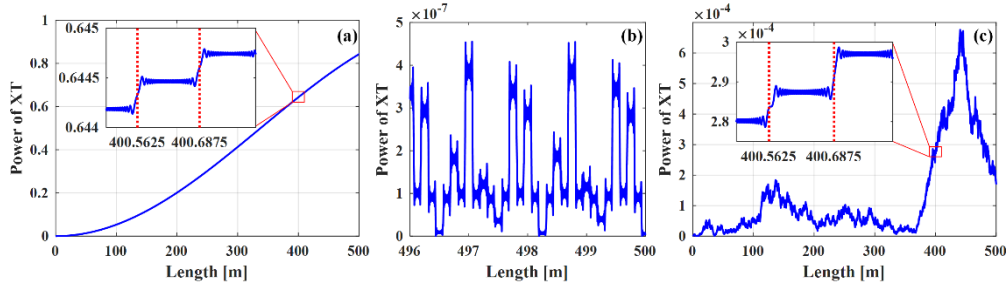

Fig. 4. (a) The evolution of XT under the first condition. (b) The evolution of XT under the second condition (with inherent propagation mismatch). (c) One realization of simulation under the third condition (with reasonable fluctuations added).

Fig. 4(a) shows the simulation results under the first condition. Due to the fiber bending and twist, at some points $\theta(z_k) = \pi(1+m)/2$ (where $m = 0, 1, 2, ...$ with $\theta_0 = 0$), $\beta_{eq,A}$ is equal to $\beta_{eq,B}$ which will lead the coupling pattern from Eq. (14) to Eq. (13). These points are called PMPs. As shown in Fig. 4(a), the term $\cos(\theta(z_k))$ in Eq. (3) will be zero at $z_k = \pi(1+m)\gamma/2$, such as 400.5625 m and 400.6875 m corresponding $m = 6409$ and $m = 6410$, respectively.

Fig. 4(b) shows the simulation results under the second condition, in which only the segment 496.0 m to 500.0 m is presented and other parts of fiber exhibit the same behavior. We set $Bias_\beta$ to $2.0 \times 10^{-5}$ with core pitch 41.1 um. Therefore, the threshold bending radius is about 2.05 m [28]. Therefore, the homogeneous WC-MCF operates in phase-matching region with bending radius 105 mm. The coupling power oscillates with transmission distance and the maximum coupling power is reduced to about $5\times 10^{-7}$. The reason is that the relative phase mismatch at each PMP is no longer uniformly distributed or randomly distributed when we do not add fluctuations for bending radius and twist.

Fig. 4(c) shows one realization of simulation results under the third condition. We add reasonable fluctuations to make the relative phase mismatch of each PMP randomly distributed. The XT power evolution shows randomness in Fig. 4(c) rather than fixed pattern as shown in Fig. 4(b). Therefore, the maximum XT power will much larger than $5\times 10^{-7}$. In addition, it can be observed that the XT power does not change linearly with transmission distance in one simulation realization.

To confirm the average XT power and XT distribution, we run the simulation for 6000 times. All the parameters are the same as those of Fig. 4(c). The fluctuations of each simulation realization are generated independently. In order to coincide with the results of Eq. (6), we carefully set $Bias_\beta$ to $2.0\times 10^{-5}$. The reason is the average XT power is sensitive to the difference of intrinsic effective refractive index as proposed in [33].

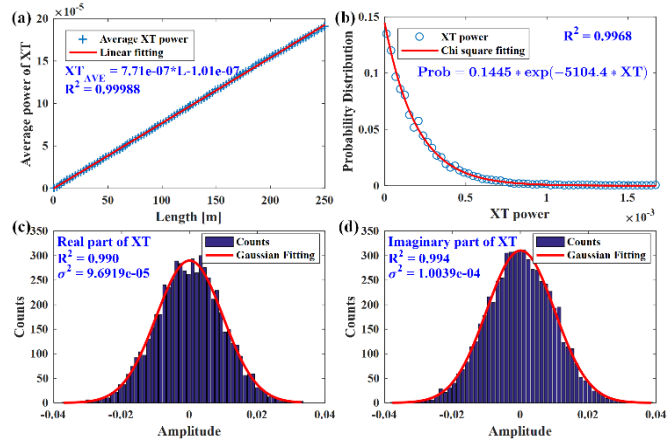

Fig. 5. (a) Average XT power. (b) Probability distribution of XT. (c) The distribution of real part of XT.

(d) The distribution of imaginary part of XT

Fig. 5(a) shows the average XT power with transmission distance. The fitting results shows the coefficient $h$ in Eq. (6) is $7.71\times 10^{-7}$. The analytical results of coefficient $h$ calculated by Eq. (6) is $7.707\times 10^{-7}$. It can be concluded that the results of the proposed simulation method will be consistent with the existing XT model [4] with reasonable parameters. In addition, we verify the statistical distribution of XT amplitude belonging to the chi-square distribution with two-degree freedom, as shown in Fig. 5 (b). We confirm that the real part and imaginary part of XT have Gaussian profiles, as shown in Fig. 5(c) and (d).

In this part, we confirmed the simulation method is suitable for analyzing XT in homogeneous WC-MCF. Since the main purpose of this work is to propose the channel model for homogeneous WC-MCF in phase-matching region, we will discuss the XT with different bending radius, twist speed, propagation constant mismatch and other parameters for homogeneous ( or heterogeneous) WC-MCF ( or SC-MCF) in our future works.

2.5. Numerical investigation of frequency-dependent XT

In order to solve the first issue of homogeneous WC-MCF's channel model, we need to figure out the evolution of XT's decorrelation bandwidth with transmission distance. Therefore we need to run the simulation with different optical frequencies but with the same simulation condition. The fluctuations are identical for all optical frequencies.

Firstly, we need to confirm the numerical simulation results under the ideal condition with analytical expression can be derived. the $Bias_\beta$, $S_A$, $S_B$, $S_{RB}$, $S_{TB}$ are all zero. The bending radius and calculation step size are 105 mm and $1.0\times10^{-4}$ m, respectively. The evolution of XT power with transmission distance has been demonstrated as Fig. 2(b) without twist and as Fig. 4(a) with twist. Fig. 6(a) show the results of frequency-dependent XT at transmission distance 400.0 m with twist (red line) or without twist (blue line). When the fiber is without twist, the optical wavelength is changing from 1549.97 nm to 1550.03 nm with 1500 sample wavelengths. When the fiber is twist, the optical wavelength is changing from 1532.00 nm to 1568.00 nm with 1500 sample wavelengths. The twist speed is 4 rounds/m. Fig.6 (b) shows the real part and imaginary part of XT without twist. It can be observed that the variation amplitudes of the real part and the imaginary part are the same.

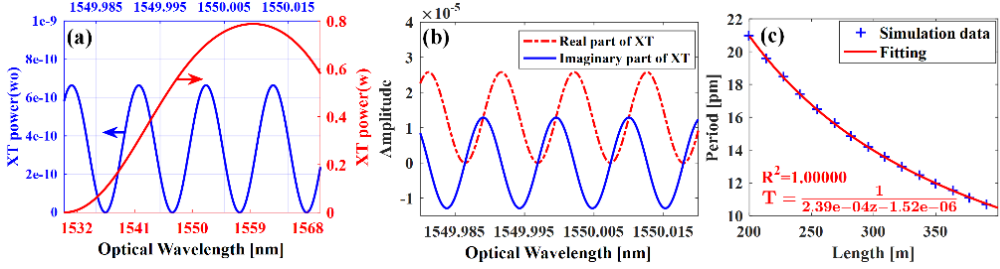

Fig. 6. (a) Power of frequency-dependent XT at transmission distance 400.0 m with twist (red line) or without twist (blue line). (b) Real and imaginary parts of XT without twist. (c) The evolution of XT's period without twist.

Since there is an analytical solution in the case of Fig.6 (b), we can theoretically analyze the evolution of the oscillation period with distance. We assume the refractive index is the same for different optical wavelengths. With this assumption, we can calculate the propagation constant $\beta_c$ and the coupling coefficient $\kappa$ with optical wavelength changing from 1545.0 nm to 1555.0 nm, as shown in Fig. 7. It can be observed that the propagation constant and the coupling coefficient changes linearly with optical wavelength.

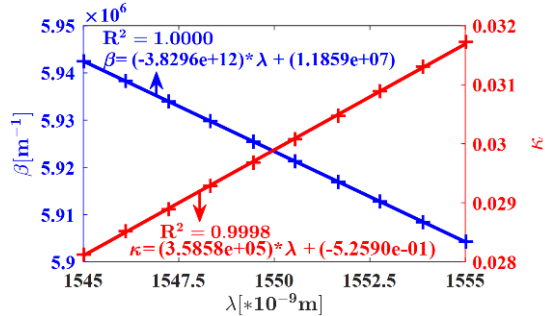

Fig. 7. Propagation constant and coupling coefficient change linearly with optical wavelength.

Therefore, the XT power can be represented as Eq. (15) derived from Eq. (14). The parameter $Q$ and $P$ are $-3.8296\times10^{12}$ and $1.1859\times10^{7}$, respectively, based on Fig. 7. The period of XT power with optical wavelength can be expressed as Eq. (16). It can be concluded that the period of frequency-dependent XT obeys the fractional linear function $1/(az+b)$ with $b=0$. The coefficient $a=2.385\times10^{-4}$ derived by Eq. (16), which is in good agreement with fitting result $2.39\times10^{-4}$ as shown in Fig. 6(c).

$$|B(z,\lambda)|^2 = \left|\frac{\kappa}{q}\sin(qz)\right|^2 \approx \left|\frac{\kappa}{q}\sin(|\delta|z)\right|^2 = \left|\frac{\kappa}{q}\sin\left(\frac{1}{2}\frac{D}{R_b}|Q\lambda+P|z\right)\right|^2 \tag{15}$$

$$T(\lambda) = \left| \frac{2\pi R_b}{QDz} \right| \tag{16}$$

In the above, we verify that the numerical simulation method can well describe the XT when there is no fluctuations and inherent propagation constant mismatch $Bias_\beta$. We point out the oscillation period of frequency-dependent XT decreases with transmission distance by fractional linear function. However, for real homogeneous WC-MCF, the fluctuations and $Bias_\beta$ should be considered. We also calculate the XT with 1500 sample wavelengths from 1544.00 nm to 1556.00 nm under the conditions shown in Fig. 5. The wavelength interval is 8.0 pm (about 1.0 GHz at 1550 nm). The fluctuations are identical for all optical frequencies.

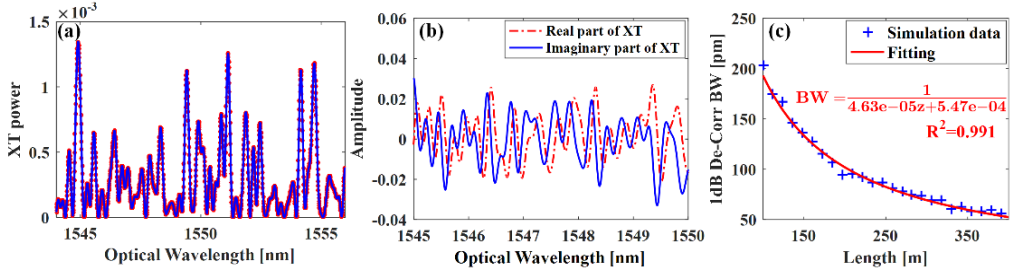

Fig. 8. (a) Power of frequency-dependent XT with fluctuations. (b) Real and imaginary parts of XT with fluctuations. (c) The evolution of XT's decorrelation bandwidth (BW).

Fig. 8(a) shows the frequency-dependent XT power changes with optical wavelength quickly and randomly at transmission distance 400.0 m, which correspond to the second construction strategy of homogeneous WC-MCF channel model. The real part and imaginary part of frequency-dependent XT are shown in Fig. 8(b) with correlation coefficient 0.0074. Therefore, they can be treated as independent with each other. Since the XT changes randomly with optical frequency, we cannot calculate the period using Eq. (16). Therefore, we calculate the decorrelation bandwidth of XT power spectrum with half width 1 dB rather than 1/e to evaluate XT power with lower variation range. Fig. 8(c) shows the evolution of decorrelation bandwidth, which is fitted well with fractional linear function $1/(az+b)$ with $R^2$ higher than 0.991.

When the homogeneous WC-MCF is in phase-matching region, the XT power changes mainly at PMPs. At other positions, the XT power shows rapid oscillation with ultra-low amplitude variation. Therefore, it is reasonable to describe that the XT happens discretely. The RPM between different cores will accumulate with transmission distance. Due to the bending, twist and other fluctuations, the RPM at each PMP is randomly distributed.

Further, the RPMs of different optical frequencies at each PMP are also different because the propagation constant will change with optical frequency. Therefore, the RPMs of different optical frequencies are different. Since the RPM at each PMP dominates the increase or decrease of XT power, the evolution of XT power of different optical frequencies at each PMP should be different, which causes the frequency-dependent characteristics of IC-XT.

In addition, the difference of RPM will increase with transmission distance, which will make the decorrelation bandwidth decrease with transmission distance. When the fiber is bending and twist and the inherent propagation constant is not zero, the RPM's differences of different optical frequencies will be approximately linearly accumulated with some fluctuations, represented as Eq. (17) based on Eq. (2), Eq. (9) and Eq. (10). The second and third integration parts Eq. (17) will not accumulated with transmission distance because of the fluctuations without bias. Therefore, the decorrelation bandwidth of XT power decreases with fractional linear function as shown in Fig. 8(c). It means if we have another homogeneous WC-MCF with other parameters, the decorrelation bandwidth will be still fitted well with fractional linear function and only the fitting parameters will be different.

$$\Delta\phi(z,\omega) = \int_0^z \beta_c(\omega) Bias_\beta dz \\ + \int_0^z \beta_c(\omega) D \frac{\cos\theta(z)}{R_b(z)}(1+Bias_\beta + S_{\beta,B})dz + \int_0^z \beta_c(\omega)(S_{\beta,B} - S_{\beta,A})dz \quad (17)$$

3. Experimental verification of frequency-dependent XT

To verify the above conclusions, we experimentally measured the frequency-dependent XT by splicing a pair of home-made fan-in/fan-out devices at both ends of the 7-core step index homogeneous WC-MCF as shown in Fig. 9 [5]. The average core pitch and the cladding diameter are 41.1 um and 150.0 um, respectively. The RI of each core and the cladding are 1.4639 and 1.4591, respectively. The WC-MCF is spooled with bending radius about 0.105 m and twisting speed about 1.0 round/m. The XT is about -11.1 dB/100 km by using Eq. (6). We used the wavelength sweeping method to get the XT power at different wavelengths [30]. The sweeping speed is set to 10.15 nm/s with wavelength ranging from 1549.5 nm to 1550.5 nm. The laser linewidth of tunable laser source (PHOENIX 1200) is less than 100 kHz. For the acquisition card, the sampling rate is set to 1 M/s with 100000 points, which means the frequency resolution is about 1.25 MHz (0.010 pm). The tunable laser and the acquisition card shared one trigger source to ensure that the wavelength does not shift at each acquisition.

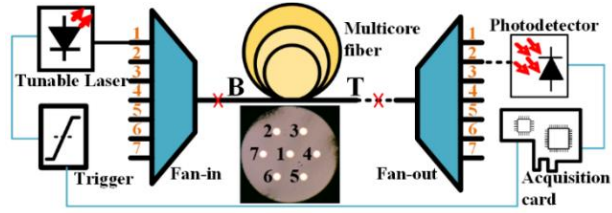

Fig. 9. Experimental scheme of XT measurement.

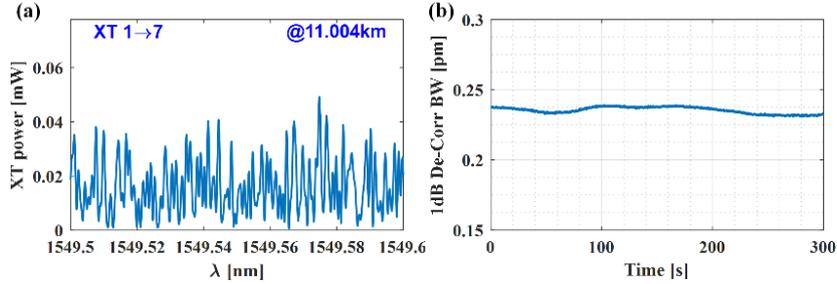

Fig. 10. (a) Frequency-dependent XT from Core 1 to Core 7 at 11.004 km. (b) Variation of 1 dB decorrelation bandwidth of XT from Core 1 to Core 7 at 11.004 km

Fig.10 (a) shows the XT from Core 1 (center core) to Core 7 under different optical frequencies with the range 12.5 GHz (0.1 nm) .The input power is about 5.0 dBm and the transmission length is 11.004 km. Moreover, we continuously measure the XT power within 300 seconds with acquisition interval about 0.65 s. The decorrelation bandwidth varies between 0.232 pm and 0.238 pm with the variation range about 1.7%. On one hand, the result shows that the estimation of decorrelation bandwidth is accurate. On the other hand, the result shows that the decorrelation bandwidth is slow time varying.

In order to measure the decorrelation bandwidth of IC-XT at different transmission lengths, we successively cut off about 500 m WC-MCF from top end and then spliced with the fan-out again. The real length of MCF is measured by optical time domain reflectometer (OTDR). When cutting off the WC-MCF, we keep the fan-in connected with bottom end of WC-MCF to keep the core label unchanged. Fig. 11 shows the XT evolution of center core (Core 1) to each outer core (Core 2-7) with different transmission lengths. All the decorrelation bandwidths are fitted with the fractional linear function $1/(az+b)$. It can be observed that the fitting parameter $a$ and $b$ are different among the six cores with all the $R^2$ are higher than 0.991.

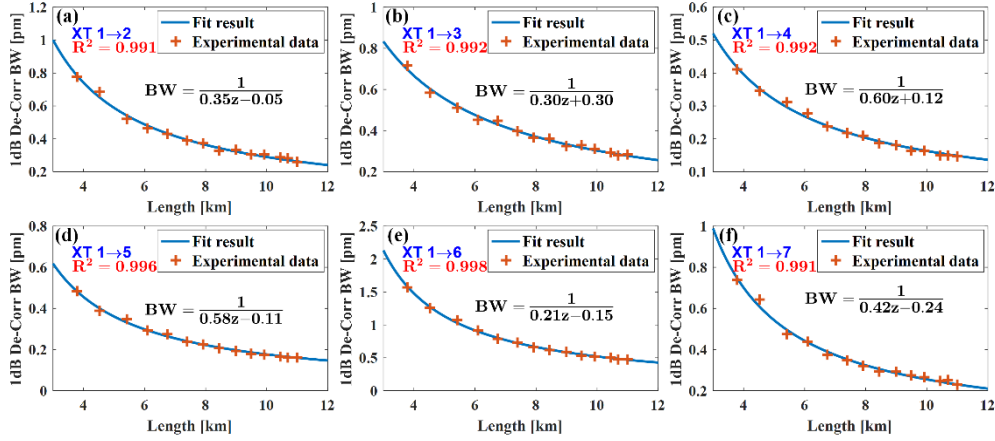

Fig. 11. The evolution of decorrelation bandwidth of XT from Core1 to Core 2-7 changed with transmission length.

4. New channel model for homogeneous WC-MCF

In the above, we solved the first issue in homogeneous WC-MCF's channel model. It can be concluded that the decorrelation bandwidth of XT power decreases with transmission distance by fractional linear function when the fiber operates in phase-matching region. In this part, we focus on the second issue that how to construct the coupling matrix with frequency-dependent characteristics automatically.

It can be observed that both the coupling coefficient $\kappa$ and propagation constant $\beta_c$ change with optical frequency shown in Fig. 7. Here, we separately assume the $\beta_c$ and $\kappa$ does not change with optical frequency to figure out which is the dominant factor inducing the frequent-dependent characteristics or both. All the other parameters are the same as Fig. 4(c). Fig. 12 shows the simulation results under three different conditions. The blue dot line represents only $\beta_c$ changes with optical frequency. The red solid line represents only $\kappa$ changes with optical frequency. It can be observed that the frequency-dependent XT is mainly dominated by the frequency-dependent propagation constant $\beta_c$. Further, the frequency-dependent propagation constant will induce the frequency-dependent RPM during transmission. Therefore, in the coupling matrix of XT, we can assume that coupling coefficient $\kappa$ does not change with optical frequency.

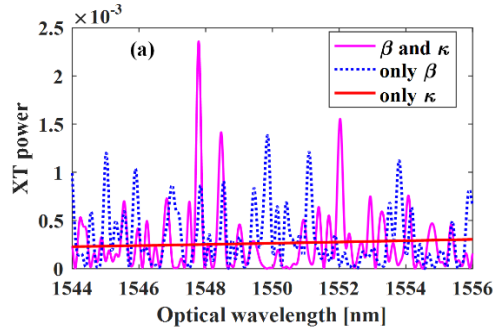

Fig. 12. Frequency-dependent XT under three different conditions (blue dot line: only the propagation constant changes; red solid line: only the coupling coefficient changes, magenta solid line: both the propagation constant and the coupling coefficient change).

To model the propagation effects in two weakly coupled cores, the NLSEs can be represented as Eq.

(18), where the symbol $A$ and $B$ is electric field amplitude of Core A and Core B, respectively. The symbol $T$, $z$, $\alpha$ and $\beta_{2,n}$ represent the time, transmission distance attenuation and GVD.

$$\frac{\partial A}{\partial z} = \left(-\frac{\alpha}{2} - \Delta\beta_1 \frac{\partial}{\partial T} - \frac{j\beta_{2,A}}{2}\frac{\partial^2}{\partial T^2}\right)A + j\gamma_A |A|^2$$
$$\frac{\partial B}{\partial z} = \left(-\frac{\alpha}{2} + \Delta\beta_1 \frac{\partial}{\partial T} - \frac{j\beta_{2,B}}{2}\frac{\partial^2}{\partial T^2}\right)A + j\gamma_B |B|^2 \tag{18}$$

$\Delta\beta_1$ describes the group velocity difference between Core A and Core B represented as Eq. (19), where the $v_{g,n}$ is the group velocity of Core n. $\gamma_n$ can be represented as Eq. (20), where $n_2$, $f_{\omega_0}$, $c$ and $A_{eff,n}$ are nonlinear index, center optical frequency, the velocity of light in vacuum and the effective area of Core n, respectively.

$$\Delta\beta_1 = \frac{\beta_{1,B} - \beta_{1,A}}{2} = \frac{1}{2}\left(\frac{1}{v_{g,B}} - \frac{1}{v_{g,A}}\right) \tag{19}$$

$$\gamma_n = \frac{2\pi n_2 f_{\omega_0}}{cA_{eff,n}} \tag{20}$$

The NLSEs (Eq. (18)) are numerically solved by split-step Fourier method independently when there is no XT between cores. When there is XT between Core A and Core B, the coupling matrix of XT can be represented in Eq. (21) based on Eq. (4). The XT is modeled as signal copies with random phase $\phi_{rnd}$ without frequency-dependent characteristics. The coupling strength $|K|$ can be calculated by Eq. (5).

$$\begin{bmatrix} A_z^{'}(z,\omega) \\ B_z^{'}(z,\omega) \end{bmatrix} = \begin{bmatrix} \sqrt{1-|K|^2} & -j|K|\exp(-j\phi_{rnd}) \\ -j|K|\exp(+j\phi_{rnd}) & \sqrt{1-|K|^2} \end{bmatrix} \begin{bmatrix} A_z(z,\omega) \\ B_z(z,\omega) \end{bmatrix} \tag{21}$$

If we take into account the frequency-dependent characteristics of XT, the coupling matrix should be rewritten as Eq. (22). We assume the coupling strength $|K|$ does not change with optical frequency because the frequency-dependent XT is dominated by the frequency-dependent RPM as mentioned above. One way to describe XT's frequency dependence is proposed in our previous work. We can calculate the XT in frequency domain. The spectrum of optical signals is manually divided into multiple frequency segments. For optical signals within one frequency segment, the IC-XT is treated as delayed copies of signals represented as Eq. (21). The random phases of different frequency segments are generated independently to simulate the IC-XT changing rapidly with optical frequency.

$$\begin{bmatrix} A_z^{'}(z,\omega) \\ B_z^{'}(z,\omega) \end{bmatrix} = \begin{bmatrix} \sqrt{1-|K|^2} & -j|K|\exp(-j\Delta\phi(z,\omega)) \\ -j|K|\exp(+j\Delta\phi(z,\omega)) & \sqrt{1-|K|^2} \end{bmatrix} \begin{bmatrix} A_z(z,\omega) \\ B_z(z,\omega) \end{bmatrix} \tag{22}$$

Here, we propose a more efficient way to construct frequency-dependent RPM automatically represented as Eq. (23). $\Delta\phi(z,\omega)$ is the RPM between Core A and Core B where $\phi_n(z,\omega)$ is represented as Eq. (24). Since the Eq. (18) does not consider the $\Delta\beta_0$ in Eq. (8), we need to add $\phi_{rnd}(\omega_0)$ to describe the RPM caused by $\Delta\beta_0$, which is assumed uniformly distributed within $[0, 2\pi]$. Since $\phi_n(z,\omega)$ records the relative phase shift of Core n compared with center optical frequency $\omega_0$ within one calculation step size $h_{xt}$, we call it phase transfer function (PTF). The symbol $angle$ represents the phase of electric field amplitude at a certain position, which can be easily calculated when the split-step Fourier method is used. It should be noted that we need to unwrap the phase before calculating PTF. As long as the fiber parameters and calculation step size are determined, the frequency-dependent RPM can be automatically calculated based on PTF.

$$\Delta\phi(z,\omega) = \phi_B(z,\omega) - \phi_A(z,\omega) + \phi_{rnd}(\omega_0) \quad (23)$$

$$\begin{aligned}\phi_A(z+h_{xt},\omega) &= angle(A(z+h_{xt},\omega)) - angle(A(z,\omega)) \\ \phi_B(z+h_{xt},\omega) &= angle(B(z+h_{xt},\omega)) - angle(B(z,\omega))\end{aligned} \quad (24)$$

In order to verify the proposed channel model, we get the evolution of XT's decorrelation bandwidth calculated by the proposed channel model. We inject an optical pulse into Core A with amplitude and initial phase 1.0 and $\pi/4$, respectively. The optical pulse is up-sampled by 8 times. The real and imaginary parts of the optical pulse are the same shown as Fig. 13(a). The spectrum of the optical pulse is shown in Fig. 13(b) with bandwidth 50 GHz.

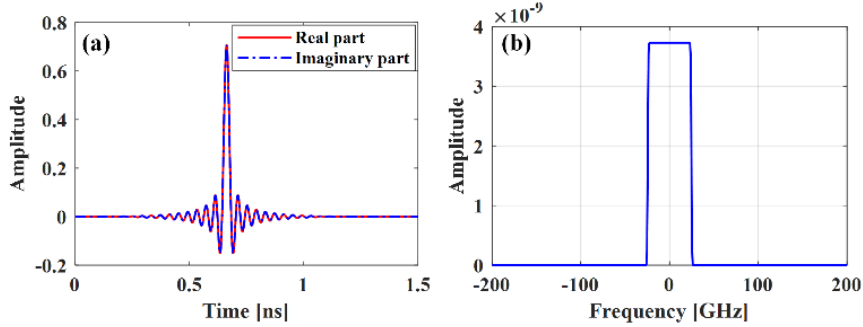

Fig. 13. (a) The real part and imaginary part of the optical pulse. (b) The spectrum of the optical pulse.

To verify the crosstalk power, we set the attenuation coefficient $\alpha$ to zero. To avoid nonlinear effect, the nonlinear coefficient $n_2$ is also zero. The center optical wavelength, the group velocity's difference and each core's dispersion are 1550 nm, 0.2 ps/m and 16.7 ps/nm/km, respectively. The other parameters of the homogeneous WC-MCF is unchanged. Eq. (18) is solved by split-step Fourier method with step size 2.0 m. The step size of XT's calculation is also 2.0 m which represents the twist speed about one round per four meters. The transmission distance is 10.0 km which means the average XT power is -25.63 dB/km calculated by Eq. (6). Fig 14(a) and (b) show the XT power spectrum and the real and imaginary part of XT at 10.0 km, respectively. The calculated XT is -25.30 dB with error less than 0.3 dB. Further, we calculate the decorrelation bandwidth at different transmission distances. The decorrelation bandwidth and fitting results are shown in Fig. 14(c) with $R^2$ higher than 0.991. Therefore, the XT's frequency-dependent characteristics of the proposed homogeneous WC-MCF's channel model are confirmed.

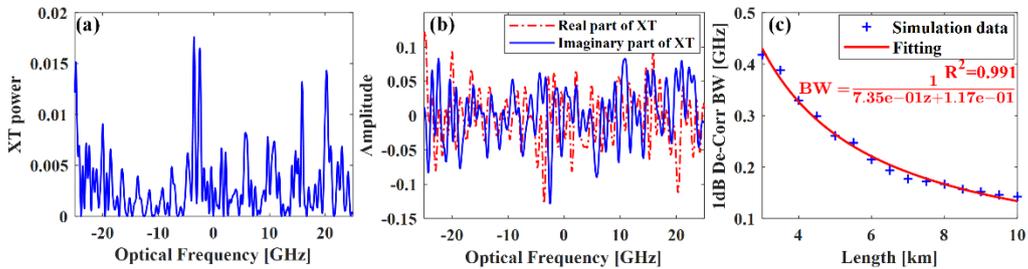

Fig. 14. The simulation results of proposed WC-MCF's channel model: (a) The power of frequency-dependent XT. (b) The real and imaginary parts of XT. (c) The evolution of decorrelation bandwidth (BW).

5. Conclusions

Two issues in homogeneous WC-MCF's channel model have been investigated. For the first issue, it can be concluded that the decorrelation bandwidth of XT decreases with transmission distance by fractional linear function. In addition, The proposed numerical simulation method for solving the modified MCEs is suitable for modeling the XT in homogeneous or heterogeneous WC- or SC- MCF. For the second issue, a new channel model has been proposed based on PTF, which can describe the frequency-dependent XT automatically. The group velocity difference, group velocity dispersion, and nonlinear propagation effects will affect the phase significantly, which is recorded in PTF. Therefore, the proposed channel model is suitable for analyzing the interactions between XT and the above linear and nonlinear transmission effects. The reason for the differences of decorrelation bandwidth between different cores should be further investigated. Without doubt, the channel model for homogeneous WC-MCF should be further verified by transmission experiments.

6. Funding

National 863 High-tech R&D Program of China under Grant (2015AA016904); National Natural Science Foundation of China under Grant (61722108, 61331010, 61205063, 61290311).